\documentstyle[pra,aps,preprint]{revtex}
\begin{document}
\draft
\title{Low voltage transport through a 
tunneling barrier in Tomonaga-Luttinger 
liquid constriction. }
\author{V.V. Ponomarenko}
\address{Department of Applied Physics, University of Tokyo,
Bunkyo-ku, Tokyo 113, Japan \\
Permanent address: A.F.Ioffe Physical Technical Institute,\\
194021, St. Petersburg, Russia}
\date{\today}
\maketitle
\begin{abstract}
As voltage decreases d.c. condctivity of a Tomonaga-Luttinger liquid
wire collapses to a small value determined by the length of the wire
and its contacts with the leads. In condition that voltage drop $(V)$
mostly occurs across a tunnel barrier inside the wire the tunneling 
density of states and, hence, the differential conductivity are shown to
exhibit an interference structure resulted from the transition of the 
Luttinger liquid quasiparticles into free electrons at the exits from
the wire. The finite length correction to the scale-invariant 
$V^{2/g-2}$ dependence of the conductivity  oscillates as a function of
voltage with periodicities related to both rigth and left traversal times.

\end{abstract}
\pacs{72.10.Bg, 72.15.-v, 73.20.Dx}

\narrowtext
Recently quantum transport in Tomonaga-Luttinger (TL) liquids has
attracted a great deal of interest \cite{phystod} as it was 
suggested to model both transport through a 1D constriction 
\cite{ar,kf} 
and the edge state transport in the fractional quantum Hall 
regime \cite{fqh}. The problem most often discussed
in this context was a suppression of the TL transport by
a point scatterer \cite{kf,pc}. Finally, its exact solution 
has been constructed \cite{es} which, however, could not be 
addressed to the 1D transport to full extent
because of the importance of 
the finite length effect in this case\cite{of,tar}. 
To handle the latter effect on the transport between
two leads through a 1D wire  the model of the ihomogeneous
TL liquid (ITL) has been invented \cite{mas,pon,safi}. It has allowed
in particular to predict an interference structure in the 
density of states of the clean wire at low energy \cite{naz}.

The aim of my present work is to examine the finite length effect
on the low voltage conductivity suppression in the ITL model of the 
one channel wire with a point scatterer inside. Temperature 
dependence of the conductance in this model was considered in 
\cite{fn}.
In spite of  simplicity the model
could be relevant to experiment due to its local stability 
against the influence of other impurities. Indeed, suppression of 
the current by an impurity is strengthened in $(t_L E_F)^{g-1}$ times 
in the TL model due to interaction. 
( Here $t_L$ is the time of travelling from
the impurity to the closest end of the wire, $E_F$ is the Fermi energy
in the wire and $g<1$ is the constant of forward scattering.) 
It means that the effect of the impurities located closer
to the middle of the wire is more important. On the other hand, 
among a few closely located impurities only one could bring about the 
strong suppression since in that case it cuts the wire and the 
renormalization of the scattering strength of the other impurities would 
be removed by the short time of travelling to this new end of the wire.
Assuming below that one impurity creates a weak link between
two parts of the wire and neglecting effects of other impurities 
one can apply the tunneling hamiltonian approach to describe the
transport.
In this approach current flowing 
through the weak link located, say, at $x=0$ inside the wire 
is given by the operator 
$J(t)=-i [T \psi^+_R(0,t) \psi_L(0,t)-h.c.]$, $(e,\hbar=1)$,
 where $T$ is the tunneling amplitude
and $\psi_{R,L}(x,t)$ are the electron annihilation operators in 
the right $0 \leq x <L_R$ and in the left $-L_L<x \leq 0$ part of the 
wire,respectively. Then the average current 
under voltage $V$ applied to the left lead is equal
in the lowest order in $T$ to:
$<J>=2 \pi |T|^2 \int d \epsilon [f(\epsilon-V)-f(\epsilon)] \rho_R(\epsilon)
\rho_L(\epsilon-V)$, 
  where $f$ is Fermi distribution. The tunneling density
of the right (left) end of the junction $\rho_{R(L)}(\epsilon)$
being the sum of the particle and hole densities can be extracted from 
the particle correlator as 
$(1-f(\epsilon)) \rho(\epsilon)=1/(2 \pi) \int d t e^{i \epsilon t}
<\psi(0,t) \psi^+(0,0)>$ or from the hole one as 
$f(\epsilon) \rho(\epsilon)=1/(2 \pi) \int d t e^{i \epsilon t}
<\psi^+(0,0) \psi(0,t)> $. 

To calculate the tunneling density to the right from the weak link
 $\rho_{R}$ let me first consider spinless fermions and apply
bosonization to the $\psi$ field under condition of an elastic reflection 
from the boundary located at $x=0$. ( Carrying out this calculations
 I will omit index "R" below. )
Above boundary condition known as
"fixed" \cite{f} in conformal field theory 
was sometimes addressed as "open" \cite{o} in other considerations. 
Bosonic repersentation  of the $\psi$ field reads 
$\psi(x,t)=\sum_{a=r,l} \psi_{a}(x,t)=(2 \pi \alpha )^{-1} 
\sum_\pm exp\{i(\theta(x,t) \pm \phi(x,t))/2 \}$, where $\psi_{r(l)}$ is
the right (left) going chiral component of $\psi$ and the 
$\theta$ and $\phi$ fields are bosonic and mutually conjugated 
$[\theta(x,t),\phi(y,t)]=2 \pi i sgn(x-y)$. 
The elastic reflection means that $\psi_{l}(0,t)= e^{i\delta }\psi_{r}(0,t)$
with an appropiate phase shift $\delta $. This results in both:
\begin{equation}
\phi(0,t)=\delta, \;\; \frac{1}{2\pi} \partial_x \theta(x,t)|_{x=0}=
\psi^+_r(0,t)\psi_{r}(0,t)-\psi^+_{l}(0,t)\psi_{l}(0,t)=0.
\label{1}
\end{equation}
Then the density of particle states could be found from
\begin{equation}
\rho_p(\epsilon)=1/(2 \pi) \int d t e^{i \epsilon t}
<\psi(0,t) \psi^+(0,0)>=\frac{\rho_O E_F}{2 \pi} 
 \int^{+\infty}_{-\infty} d t e^{i \epsilon t +\frac{1}{4}
[<\theta(0,t)\theta(0,0)>-<\theta^2(0,0)>]}
\label{2}
\end{equation}
where the value of the free electron tuneling density was 
introduced as: $\rho_O=(1+\cos(2 \delta ))/(\pi v)$.

The problem reduces to finding the $\theta$ field correlator. It 
can be done for the finite length piece of the wire adiabatically 
connected to the lead making use of the ITTL model \cite{mas,pon,safi}. 
In this model
the Tomonaga-Luttinger interaction $(\sum_{r,l}\rho_a)^2$ is switched
on in the wire $x<L_R$ and switched off outside. Then the Hamiltonian
takes a bosonized form
\begin{equation}
{\cal H}= \int_0^{\infty} dx  \frac{v}{2}
\{u^2(x)  
\left({{\partial_x \phi(x) } \over {\sqrt{4 \pi}}} \right)^2 + 
\left({{\partial_x \theta(x) } \over {\sqrt{4 \pi}}}
\right)^2 \}
\label{3}
\end{equation}
where function $u(x)$ ensuing from the interaction can be approximated
in the low energy limit by a step-function: $u(x)=1$ if $x>L_R$ and
$u(x)=u=1/g<1$, otherwise. The correlator of the $\theta$ field ordered
in imaginary time $T(x,y,\tau)\equiv <T_{\tau}{\theta(x,\tau)
\theta(y,0)}>$ can be shown to 
satisfy the following equation 
\begin{equation}
\{\frac{1}{v^2 u^2(x)} \partial^2_\tau + \partial^2_x \} T(x,y,\tau)=
-\frac{4 \pi}{v} \delta(x-y)\delta(\tau)
\label{4}
\end{equation}
under the boundary conditions $\partial_x T(x,y,\tau)|_{x=0}=0$ 
following from (\ref{1}). 
Fourier transform of this correlator $T(x,y,\omega)$
is symmetrical under $\omega\rightarrow -\omega$. It can be 
compiled from the solutions of the homogeneous equation corresponding
to Eq.(\ref{4})
\begin{eqnarray}
\{\frac{\omega^2}{v^2 u^2(x)} - \partial^2_x \} f_\omega(x)=
\{\frac{\omega^2}{v^2 u^2(x)} - \partial^2_x \} h_\omega(x)=
\frac{4 \pi}{v} \delta(x-y)
\label{5}\\
T(x,y,\omega)=\frac{4 \pi}{v W(\omega)}
[\theta(x-y)f_\omega (x) h_\omega (y)+\theta(y-x)f_\omega (y) h_\omega (x)]
\nonumber 
\end{eqnarray}
if these solutions meet  boundary conditions:
$h'_\omega(0)=0,  f_\omega (x)=exp(-\omega x/v)$ at 
$x \rightarrow \infty$ and positive $\omega$. 
The Wronskian $W(\omega)$ is equal to $-f'_\omega (0) h_\omega(0)$
and, hence, $T(0,0,\omega)=-\frac{4 \pi}{v} 1/(ln f_\omega (0))'$. 
The only solution I need can be written as right going plus
reflected left going waves at $x<L_R$. The reflection amplitude 
$r=-e^{-2 \eta},\  \tanh(\eta)=1/u$ equals minus reflection amplitude
of the $\phi$ field \cite{safi,pon2} due to duality symmetry and
is negative for the repulsive interaction. Substituting this solution
one can find 
$T(0,0,\omega)=\frac{4 \pi u}{\omega} \tanh(\omega t_{LR} + \eta)$ with
$t_{LR}$ equal to the time of
travelling  from the junction to the right lead. 
Analytical continuation of this function $[-T(0,0,-i\omega+0)]$ brings 
us the value
of the retarded Green function for the $\theta$ field. Imaginary 
part of
the latter multiplied by the Bose distribution function for holes
$1+f_B(\omega)$ and by a factor $(-2)$
coincides with the Fourier transform of the correlator at $\omega$ what
allows me to rewrite expression for the particle density of states
(\ref{2}) in dimensionless units as
\begin{equation}
\rho_p(\varepsilon)=\frac{\rho_O}{2 \pi \gamma }
 \int^{+\infty}_{-\infty}
 d p exp \{i \varepsilon p +u \int^\infty_\infty d \omega
e^{-\gamma |\omega |}
(1+f_B(\omega))\frac{e^{-i\omega p}-1}{\omega} 
\frac{Im \tan(\omega+i \eta)}{\tanh(\eta )}\}
\label{6}
\end{equation}
where the temperature and energy were scaled as 
$\varepsilon=t_{LR} \epsilon$. 
The dimensionless cut-off parameter $\gamma$ equals
$(E_F t_{LR})^{-1}$.
The hole density of states $\rho_h(\epsilon)$ can be
found as $\rho_h(\epsilon)=\rho_p(-\epsilon)$.  
Below I will examine expression (\ref{6})
at zero temperature when the expected effect of the interference on the 
density of states should be the most profound. The bosonic distribution
factor $1+f_B$ restricts the integral to positive $\omega$ and the density
of states at positive $\varepsilon $ coincides with the density of particle
states. 

The form of the correlator used in (\ref{6}) is equivalent to
\begin{equation}
<\psi_r(0,t)\psi^+_{r}(0,p)>=\frac{E_F}{2 \pi v} \prod _n 
\left(\frac{\gamma+2in}{\gamma+2in+p} \right) ^{ur^n}
\label{7}
\end{equation}
where $p$ is dimensionless time $p=t/t_{LR}$. This expression can be
easily comprehended as if it is a product of
the contributions of the $2n$ length paths connecting $(0,p)$ and $(0,0)$
points and undergoing $n$ reflections 
from a $x=L_R$ non-elastic boundary
with the negative reflection amplitude $r=-e^{-2 \eta}$ 
and $n$ reflections 
from the $x=0$ elastic boundary
with unit reflection amplitude.
Another form of the exponent in the
left hand side of Eq. (\ref{6}) 
$exp\{u(S(ip+\gamma)+\int_0^\infty dz/(2z) [e^{-zp}-1][\tanh(z+\eta)-
\tanh(z-\eta)])\}$
can be obtained after rotation of the integration contour $\omega=-iz$.
Here $S(ip+\gamma)=-i\pi \int_0^p ds e^{-(\eta+i\pi/2)(s-i\gamma)}
(1-e^{-i\pi(s-i\gamma)})^{-1}$ ensues from the pole contributions. For a small
$\eta$ it gives expression for the correlator
\begin{equation}
<\psi_r(0,t)\psi^+_{r}(0,p)>=\frac{E_F}{2 \pi v}
\frac{(\pi \gamma/4)^u}{[\tanh(\pi(\gamma+ip)/4)]^{ue^{-\eta p}}} 
e^{u/2\int_0^\infty (dz/z) [e^{-zp}-1][\tanh(z+\eta)-
\tanh(z-\eta)])}
\label{8}
\end{equation}
where the divergent at $p=4n$ part and the smooth longly decaying tail
finally approaching $const/p$ at $p\eta\gg 1$ are separated. Such a 
long time decay brings about the finite $\rho(0)$ value predicted before
\cite{mas2,of,ar}.

Besides making use of the above form of the correlator in (\ref{6})
it is convenient to represent the density as a sum 
$\rho(\varepsilon)/\rho_O=\frac{\gamma^{u-1}}{\pi}( \sin(\pi u)
\Gamma(1-u) \varepsilon^{u-1} + 2 r(\varepsilon))$ 
of a scale 
invariant density of the infinite long wire and the finite length
correction. Then calculation of the latter reduces to Fourier
transform of the integrable function  at least if $u<3$. It has 
been done numerically and results are depicted in Fig.1a. They
show that as the interaction is strengthened the dimensionless finite
length correction to the density of states changes its period of
oscillations from $\pi/t_{LR}$ for $u=1.4$ to $\pi/(2t_{LR})$ for $u=1.86$.
At larger $u$ the oscillations become more profound and practically
cease to decay with increase of the energy. As $u$ decreases to 1
the finite length correction $r(\varepsilon)$ goes to zero everywhere
except for $\varepsilon=0$ where the value $r(0)$ approaches $\pi/2$.
Albeit the amplitude of the $r$ function oscillations becomes much
smaller than 1 in this limit it decreases slowly with increase of the 
energy.

To generalize  expression (\ref{6}) for the density of states
to the case of the spin electrons one
can notice that the $\theta_{\sigma}$ arising in Eq. (\ref{2}) in this case
may be represented as a sum over the charge and spin fields
$\theta_{\sigma}=(\theta_{c} \pm \theta_{s})/\sqrt{2}$. Dynamics of the
charge field is described by the Hamiltonian (\ref{3}) as before while
dynamics of the spin field is not affected by the interaction. It means
that the spin generalization of the density of states requires
change of $Im[\tan(\omega+i \eta)]/\tanh(\eta )$ in (\ref{6}) into 
$(1+Im[\tan(\omega+i \eta)]/\tanh(\eta ))/2$. The results of calculations
for the spin electrons are shown in Fig.1b. They behave similar to the
results of the spinless case, however, the interference structure
is weaker than in the spin case at the same value of the interaction
constant.

Calculating the differential conductivity one can use the same form
where the finite length correction is separated from the scale invariant 
infinite length contribution:
$\partial J/ \partial V=R_O^{-1} \gamma^{2(u-1)}[(2u-1)\sqrt{\pi}
2^{1-2u}(\Gamma(u)\Gamma(1/2+u))^{-1}v^{2(u-1)}+\partial j(v)]$.
Here $R_O^{-1}$ is a free electron conductance of the junction and both the 
cut-off $\gamma$ and the applied voltage $V$ are measured in the unit of
the full traversal time $t_L=t_{LR}+t_{LL}$ so that $v=V t_L$. The finite
length correction $\partial j$ to the conductivity depends on relation 
$t_{LR}/t_{LL}$ between the travelings times to the right and left leads.
Its behavior is illustrated by Fig.2a,b. They show that as the voltage 
increases the correction grows up if $u$ is larger than $u\approx  2$ and
just oscillates otherwise. It is reasonable since the leading contribution
to the correction comes from convolution of the oscillating $r(\varepsilon)$
function and the infinite length $\varepsilon^{u-1}$ function. The amplitude
of the oscillations behaves similar to the one of the density of states.
It is on the order of the conductance for large $u$ and smaller than  
conductance for $u$ close to 1. However, in both cases the oscillations
decay very slowly as the voltage increases what hopefully makes them
observable.

In summary, I have shown that the ITLL model accounting for the finite
length of the wire and predicting a finite zero temperature conductance 
also brings about the oscillating interference structure in the 
differential conductivity. The latter survives at energies much higher
than the one corresponding to the wire length $1/t_L$. The period of the
oscillations of the tunneling density of states corresponds to the 
right (left) traversal time for a weak interaction $ u \approx 1$ and
becomes two times shorter at larger $u$. The differential conductivity
in turn oscillates with both periods of the right and left tunneling
density of states.

The author acknowledges N. Nagaosa, S. Tarucha for useful discussions.
It is a special pleasure to thank T. Iitaka for his help in conducting
calculations.
This work was supported by the Center of Excellence  and 
partially by the fund for the development of collaboration
between the former Soviet Union and Japan at the JSPS.

\figure{Dependences of the finite length correction 
$(E_F t_L)^{1-u}2r(E)/\pi$ to the
density of states on the energy $E$ measured in $\pi$ over traversal time
$t_L$ unit . In the spinless case (a) solid, dot dashed and
dashed lines correspond to constant of interaction
$u$ equal to $u=1.396,1.862,2.825$, appropriately.
In the spin case (b) solid and dashed lines correspond to 
$u=1.862,2.825$. Long decay of the oscilations at u=1.396 is shown 
in the inset.}
\figure{Dependences of the finite length correction 
$(E_F t_L)^{2(1-u)} \partial j(v)/R_O$ to the differential conductivity
on volage $v$ measured in $\pi$ over traversal time $t_L$ unit. 
For spin electrons (a) $u=1.396$ the solid line relates to the symmetrical
position of the weak link, the dotted line to the case when relation between
the lengths of the right and left shoulders is 1/4. In the spinless case
(b) the lines 1,2 correspond to $u=2.164,1.862$, respectively.}


\begin{references}

\bibitem{phystod} B.G.Levy, Physics Today, June,  21 (1994).

\bibitem{ar} W.Apel and T.M.Rice, Phys. Rev. B {\bf 26}, 7063 (1982).

\bibitem{kf}C. L. Kane and M. P. A. Fisher, 
Phys. Rev. Lett. {\bf 68}, 1220 (1992); Phys. Rev. B {\bf 46}, 15233 (1992).

\bibitem{fqh}X.G.Wen, Phys.Rev. B {\bf 40}, 7387 (1989).

\bibitem{pc}K.Moon {\it et al }, Phys. Rev. Lett. {\bf 71}, 4386 (1993);
K. A. Matveev, D. Yue, and L. I. Glazman, Phys. Rev. Lett. {\bf 71}, 3351 (1993); 
F. Guinea  {\it et al }, Europhys. Lett.  {\bf 30}, 561 (1995).

\bibitem{es}P.Fendley, A.W.W.Ludwig and H.Saleur, Phys. Rev. Lett. {\bf 75},
2196 (1995), Phys. Rev. B {\bf 52}, 8934 (1995).

\bibitem{of}M. Ogata and H. Fukuyama, Phys. Rev. Lett. {\bf 73},
468 (1994).

\bibitem{tar}S. Tarucha, T. Honda, and T. Saku,  Solid State Commun.
{\bf 94}, 413 (1995).


\bibitem{mas}D. L. Maslov and M. Stone, Phys. Rev. B {\bf 52},
R5539 (1995).

\bibitem{pon}V. V. Ponomarenko, Phys. Rev. B {\bf 52},
R8666 (1995).

\bibitem{safi}I. Safi and H. J. Schulz,
Phys. Rev. B {\bf 52}, R17040 (1995).


\bibitem{naz}Y.V.Nazarov, A.A.Odintsov and D.V.Averin, unpublished.

\bibitem{fn}A.Furusaki and N.Nagaosa,
Phys. Rev. B {\bf 54}, R5239 (1996).

\bibitem{f}.M.Ameduri, R.Konik and A.LeClair, hep-th/9503088;
H.W.J.Blote,J.L.Cardy and M.P.Nightingale, Phys. Rev. Lett. {\bf 56},
742 (1986).
\bibitem{o}.M.Fabrizio and A.O.Gogolin, Phys. Rev. B {\bf 51}, 
17827 (1995);Y. Wang, J. Voit and Fu-Cho Pu, cond-mat/9602086;
S. Eggert {\it et al }, cond-mat/9511046.


\bibitem{pon2}V. V. Ponomarenko, Phys. Rev. B {\bf 54},
10328 (1996).

\bibitem{mas2}D. L. Maslov, Phys. Rev. B {\bf 52},
R14368 (1995).


\end{references}
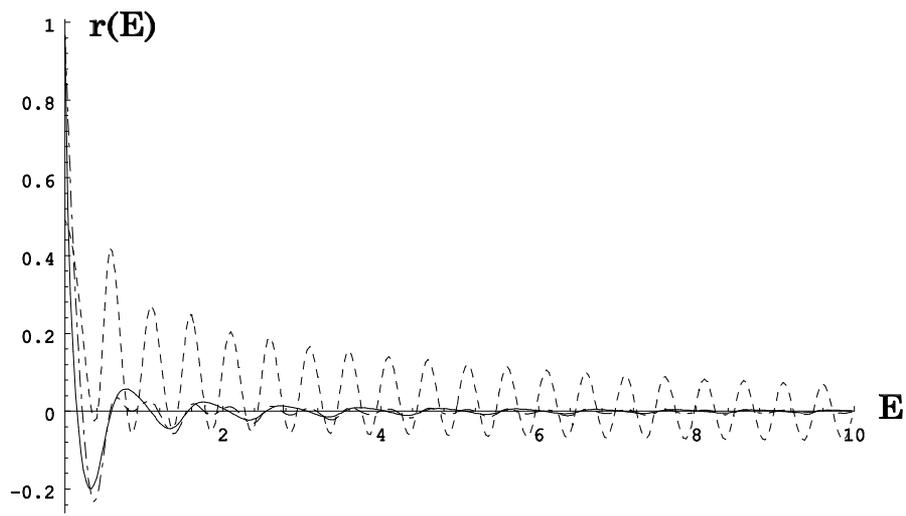
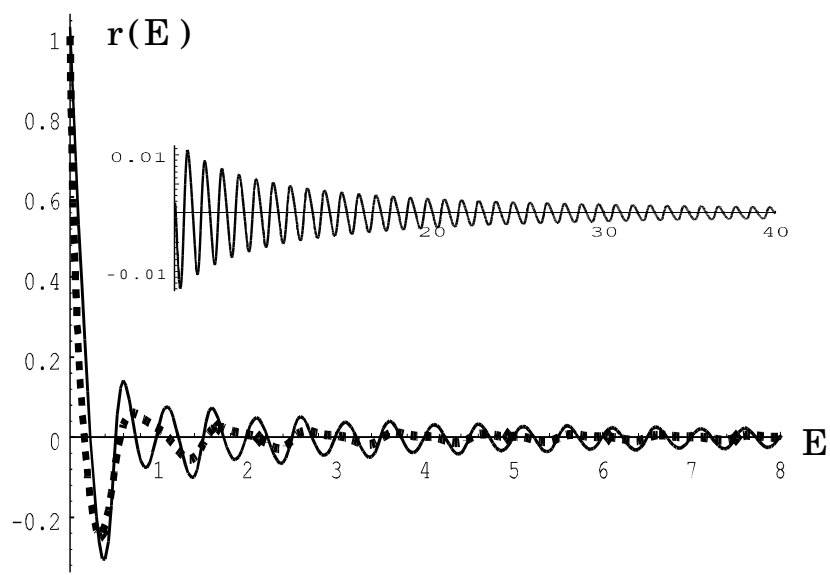
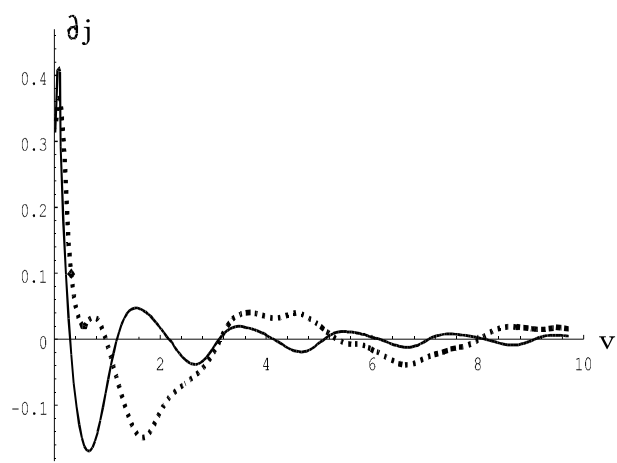
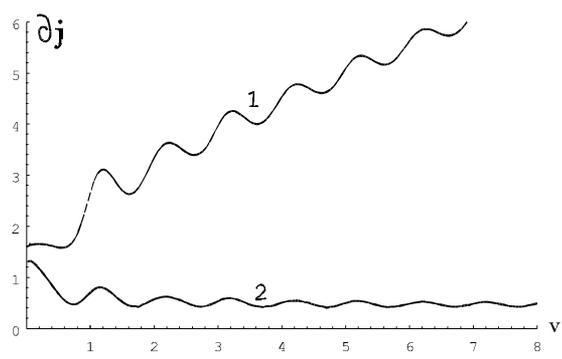
\end{document}